\documentclass[12pt]{article}
\newlength{\absize}
\pdfoutput=1
\usepackage{latexsym}
\usepackage{amssymb}
\usepackage{graphicx}
\usepackage{footmisc}
\usepackage{braket}
\renewcommand{\arraystretch}{2}
\newcommand{\figsize}{\small}

\usepackage{pictexwd}
\newdimen\tdim
\tdim=\unitlength
\def\stpltsmbl{\setplotsymbol ({\small .})}

\newbox\sru
\setbox\sru=\hbox{\beginpicture
\setcoordinatesystem units <\tdim,\tdim>
\stpltsmbl
\setquadratic
\plot
  0.0   0.0
  4.8   1.5
  7.5   5.0
  7.3   8.5
  5.0  10.0
  2.7   8.5
  2.5   5.0
  5.2   1.5
 10.0   0.0
/
\endpicture}
\def\springru #1 #2 *#3 /{\multiput {\copy\sru}  at
#1 #2 *#3 10 0 /}

\newcommand{\cl}[1]{\mathcal{#1}}
\newcommand{\clog}{\mathop{\rm Log}}

\usepackage{hyperref}

\begin{document}

\thispagestyle{empty}
\pagestyle{empty}
\renewcommand{\thefootnote}{\fnsymbol{footnote}}
\newcommand{\starttext}{\newpage\normalsize
 \pagestyle{plain}
 \setlength{\baselineskip}{3ex}\par
 \setcounter{footnote}{0}
 \renewcommand{\thefootnote}{\arabic{footnote}}
 }
\newcommand{\preprint}[1]{\begin{flushright}
 \setlength{\baselineskip}{3ex}#1\end{flushright}}
\renewcommand{\title}[1]{\begin{center}\LARGE
 #1\end{center}\par}
\renewcommand{\author}[1]{\vspace{2ex}{\Large\begin{center}
 \setlength{\baselineskip}{3ex}#1\par\end{center}}}
\renewcommand{\thanks}[1]{\footnote{#1}}
\renewcommand{\abstract}[1]{\vspace{2ex}\normalsize\begin{center}
 \centerline{\bf Abstract}\par\vspace{2ex}\parbox{\absize}{#1
 \setlength{\baselineskip}{2.5ex}\par}
 \end{center}}

\title{Physics Fun with Discrete Scale Invariance}
\author{
 Howard~Georgi,\thanks{\noindent \tt hgeorgi@fas.harvard.edu}
 \\ \medskip
 Center for the Fundamental Laws of Nature\\
 The Physics Laboratories \\
 Harvard University \\
 Cambridge, MA 02138
 }
\centerline{06/16}
\abstract{I construct a quantum field theory model with discrete scale invariance
at tree level. The model has some unusual mathematical properties (such as
the appearance of $q$-hypergeometric series) and may possibly have some
interesting physical properties as well.  In this note, I explore some
possible physics that could be regarded as a violation of standard
effective field theory ideas.} 

\starttext

\setcounter{equation}{0}

\section{Introduction\label{intro}}

In this note, we discuss a simple (though infinitely large) quantum field theory that
has a formal discrete scale invariance in tree approximation.  The original
motivation (probably not successfully realized) 
was to get some insight into theories in which the standard model
communicates with a conformal sector. But this construction leads to
interesting mathematics 
and perhaps to some interesting physics.  
A wildly optimistic hope is that studying this
simple model might yield insights into the hierarchy puzzle. In this note,
we will focus on 
the physics, and relegate some amusing mathematical results to \cite{math}.

The finite version of our model has been considered before (see for example
\cite{Cheng:2001nh,Abe:2002rj,Katz:2004qa,Falkowski:2002cm,Randall:2002qr,Randall:2005me})
as a
discretization of of a warped extra dimension to study important physical
effects like the running of couplings.  We hope that our explicit
results in the infinte theory may be useful for these important applications.

In section~\ref{dsi}, we introduce the field theory 
model and make the connection with a
system of springs and masses. In section~\ref{gaugecouplings}, we analyze
the gauge couplings of the mass-eigenstate massive vector bosons. 
In section~\ref{tale}, we discuss one way of thinking about the inifinite
theory as a limit us finite theories, and we continue in
section~\ref{tail} to discuss how the approach to the limit depends on
boundary conditions. 
In section~\ref{physics}, we will suggest that 
the dependence on boundary conditions can be much larger
than predicted in a naive effective theory picture. 
{\figsize\begin{figure}[htb]
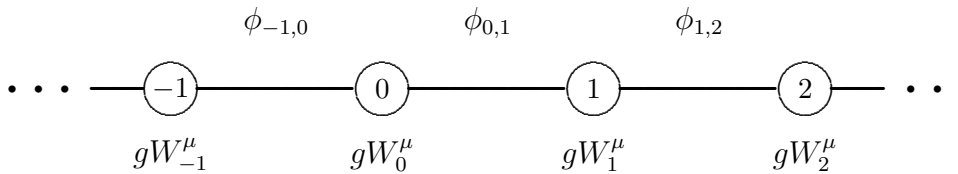

$$\beginpicture
\setcoordinatesystem units <\tdim,\tdim>
\circulararc 360 degrees from 10 0 center at 0 0
\circulararc 360 degrees from 90 0 center at 80 0
\circulararc 360 degrees from 170 0 center at 160 0
\circulararc 360 degrees from 250 0 center at 240 0
\stpltsmbl
\multiput {\tiny$\bullet$} at -60 0 *2 10 0 /
\plot -30 0 -10 0 /
\put {\small$-1$} at 0 0
\plot 10 0 70 0 /
\put {$\phi_{-1,0}$} at 40 25
\put {\small$0$} at 80 0
\plot 90 0 150 0 /
\put {$\phi_{0,1}$} at 120 25
\put {\small$1$} at 160 0
\plot 170 0 230 0 /
\put {$\phi_{1,2}$} at 200 25
\put {\small$2$} at 240 0
\plot 250 0 270 0 /
\multiput {\tiny$\bullet$} at 280 0 *2 10 0 /
\put {$gW_{-1}^\mu$} at 0 -25
\put {$gW_{0}^\mu$} at 80 -25
\put {$gW_{1}^\mu$} at 160 -25
\put {$gW_{2}^\mu$} at 240 -25
\endpicture$$
\caption{\figsize\sf\label{fig-1}A deconstructed dimension..}\end{figure}} 

\section{Discrete Scale Invariance\label{dsi}}

We start with a deconstructed
dimension as an infinite, linear
moose described in the notation of \cite{Georgi:2004iy}
by the diagram shown in figure~\ref{fig-1}.
This describes a series of $SU(N)$ groups labeled by the integers in the circles and
scalars  transforming like $\phi_{j,j+1}\sim\left(N,\bar N \right)$ 
under $SU(N)_j\times
SU(N)_{j+1}$ described by the solid lines between neighboring circles, as
shown in figure~\ref{fig-1}.  The gauge couplings are all assumed to equal
$g$ and the gauge fields are $W_j^\mu$.
{\figsize\begin{figure}[htb]
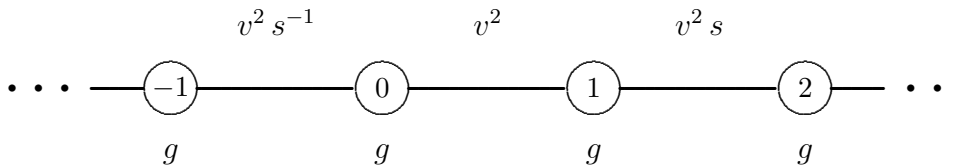

$$\beginpicture
\setcoordinatesystem units <\tdim,\tdim>
\circulararc 360 degrees from 10 0 center at 0 0
\circulararc 360 degrees from 90 0 center at 80 0
\circulararc 360 degrees from 170 0 center at 160 0
\circulararc 360 degrees from 250 0 center at 240 0
\stpltsmbl
\multiput {\tiny$\bullet$} at -60 0 *2 10 0 /
\plot -30 0 -10 0 /
\put {\small$-1$} at 0 0
\plot 10 0 70 0 /
\put {$v^2\,s^{-1}$} at 40 25
\put {\small$0$} at 80 0
\plot 90 0 150 0 /
\put {$v^2$} at 120 25
\put {\small$1$} at 160 0
\plot 170 0 230 0 /
\put {$v^2\,s$} at 200 25
\put {\small$2$} at 240 0
\plot 250 0 270 0 /
\multiput {\tiny$\bullet$} at 280 0 *2 10 0 /
\multiput {$g$} at 0 -25 *3 80 0 /
\endpicture$$
\caption{\figsize\sf\label{fig-2}A discretely scale invariant system.}\end{figure}} 
{\figsize\begin{figure}[htb]
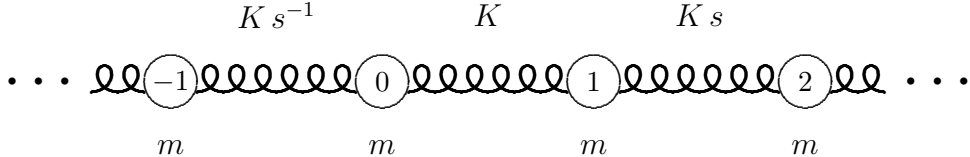

$$\beginpicture
\setcoordinatesystem units <\tdim,\tdim>
\circulararc 360 degrees from 10 0 center at 0 0
\circulararc 360 degrees from 90 0 center at 80 0
\circulararc 360 degrees from 170 0 center at 160 0
\circulararc 360 degrees from 250 0 center at 240 0
\stpltsmbl
\multiput {\tiny$\bullet$} at -60 0 *2 10 0 /
\springru -30 -4 *1 /
\put {\small$-1$} at 0 0
\springru 10 -4 *5 /
\put {$K\,s^{-1}$} at 40 25
\put {\small$0$} at 80 0
\springru 90 -4 *5 /
\put {$K$} at 120 25
\put {\small$1$} at 160 0
\springru 170 -4 *5 /
\put {$K\,s$} at 200 25
\put {\small$2$} at 240 0
\springru 250 -4 *1 /
\multiput {\tiny$\bullet$} at 280 0 *2 10 0 /
\multiput {$m$} at 0 -25 *3 80 0 /
\endpicture$$
\caption{\figsize\sf\label{fig-3}The mechanical analog of the system in
figure~\protect\ref{fig-2}.}\end{figure}} 

But the unusual
thing we will do here is to assume that
the expectation values of the fields
grow exponentially with $j$
\begin{equation}
\Braket{\phi_{j,j+1}}=v\,s^{j/2}
\label{phij}
\end{equation}
where $s$ is a positive number.
This form obviously has a discrete scale invariance under a combination of
rescaling by $s^{1/2}$ and translating one unit in $j$.  The expectation
value (\ref{phij}) can in fact arise at tree level from a scalar potential
with the same discrete scale invariance.
We don't expect this discrete scale invariance to survive beyond tree
appoximation in this model.  But I believe that 
the structure of the massive gauge bosons and couplings
may be interesting as an illustration of some general principles.  

And besides,
it is fun, because this model can be mapped onto the longitudinal
oscillations of the system of masses and
springs shown in figure~\ref{fig-3}.

The correspondance is this.  If $K\to v^2$  and $m\to1/g^2$, the angular frequencies
of the normal modes of the system $\omega_\alpha$ are proportional to the
masses of the heavy gauge bosons, $M_\alpha$.   

For $s=1$, this is just the familiar system
in which the normal modes are infinite waves and the massive vector bosons
are delocalized --- spread over all $j$ as in a deconstructed
dimension.~\cite{ArkaniHamed:2001nc}  
However (as was noted qualitatively in \cite{Abe:2002rj} for the field
theory)
unlike the situation with $s=1$, the normal
modes for $s\neq1$ are not infinitely spread out.  They go to zero for
$j\to\pm\infty$ and for very large or very small $s$, each mode   
is localized around one block.  

For large $s$, it is very plausible that there is a mode in which block $0$
oscillates with
$\omega\approx\omega_0=\sqrt{K/m}$.  It is worth thinking qualitatively
about the physics of this for large $s$, where the strength of the springs
increases as we go to the right, for increasing $j$.  For an oscillation with
$\omega\approx\omega_0$,
block 1 is nearly stuck in place by the
stiff spring to its right.  The oscillation of block $0$ is primarily due
to the spring on its right, and block 0 barely feels the effect of the weak
spring to its left.  So we expect that as $s\to\infty$, there will be a
mode in which block 0 oscillates with $\omega\to\omega_0$  while the
displacements of all the other blocks go to zero.  For large but finite
$s$, the displacements in this mode drop off away from $j=0$.  To the
right, for $j>0$, the displacements fall off exponentially.  One can think of each
spring driving the block to its right below resonance, so all the
displacments are in phase.  To the left, for $j<0$, the displacements
oscillate in sign, because the blocks are driven above resonance by the
springs to their right.

{\figsize\begin{figure}[htb]
$$\includegraphics[width=.8\hsize]{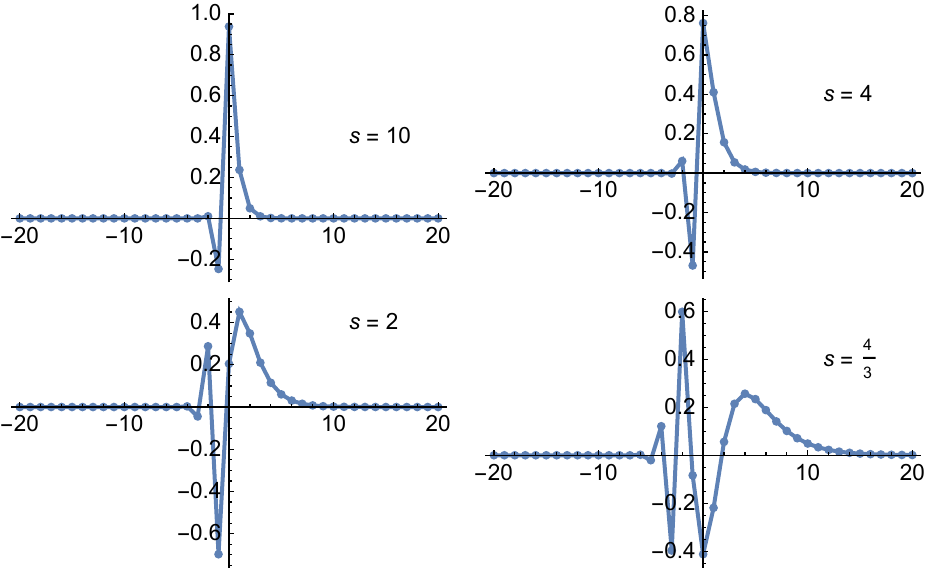}$$
\caption{\figsize\sf\label{fig-4}The amplitudes for the $\omega=\omega_0$
mode versus $j$ for various
values of $s$.}\end{figure}}
It is straightforward to turn this physical picture into a perturbative
calculation in $1/s$, and the result is a bit surprising.  The first surprise is
that $\omega$ is {\bf exactly equal} to $\omega_0$ for any $s\neq1$. 
Using the perturbative result that there is a mode with $\omega=\omega_0$,
we can write a recursion relation for the displacement $\psi_j(s)$ of the
$j$th block in this mode 
very simply as
\begin{equation}
s^j\,\Bigl(\psi_j(s)-\psi_{j+1}(s)\Bigr)
+s^{j-1}\Bigl(\psi_j(s)-\psi_{j-1}(s)\Bigr)
-\psi_j(s)=0
\label{rec1}
\end{equation}

{\figsize\begin{figure}[htb]
$$\includegraphics[width=.8\hsize]{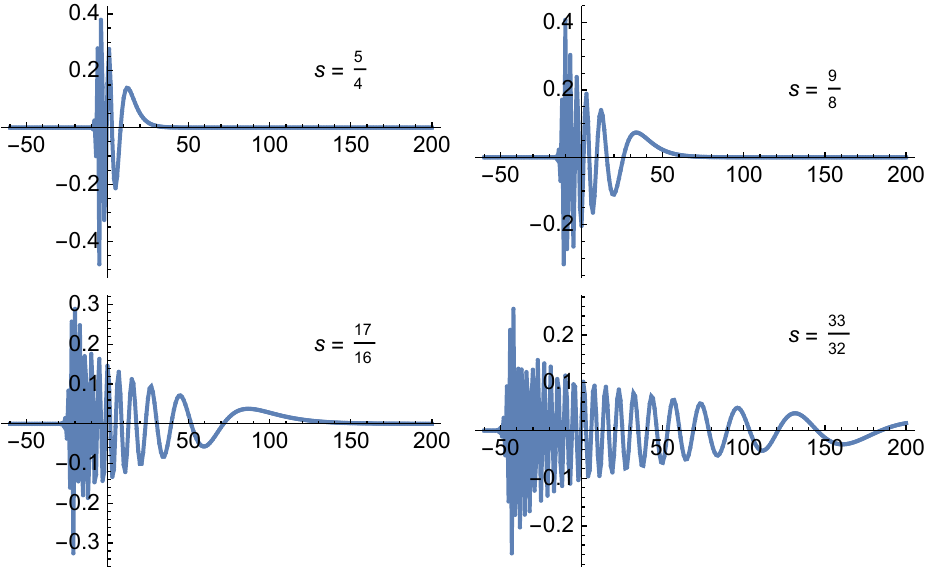}$$
\caption{\figsize\sf\label{fig-5}The amplitudes for the $\omega=\omega_0$
mode versus $j$ for
values of $s$ closer to 1.}\end{figure}}
The second
big surprise (at least to me), discussed in detail in \cite{math},
is that 
the infinite series for the mode
with $\omega=\omega_0$, 
\begin{equation}
\psi_j(s)=\sum_{\ell=1}^\infty(-1)^{\ell-1}
\,s^{(1-j)\ell}\,s^{-\ell(\ell+1)/2}\,(1-s^{-\ell})
\,
\left(\prod_{k=1}^{\ell}\frac{1}{(1-s^{-k})^2}\right)
\label{invert3i}
\end{equation}
can be summed and 
expressed in terms of $q$-hypergeometric series\footnote{I do not know
whether this is related to the $q$-Bessel functions discussed in
\protect\cite{deBlas:2006fz}.} as follows\footnote{The 
form \protect{(\ref{invert5qpatrick})} was obtained from
\protect{(\ref{invert5q})} by Patrick Komiske}

\begin{equation}
\psi_j(s) =\,
_1\phi_1\Bigl(\{0\},\{s\},s,s^{1-j}\Bigr)
\,-\,
_1\phi_1\Bigl(\{0\},\{s\},s,s^{2-j}\Bigr)
\label{invert5q}
\end{equation} 
\begin{equation}
 =-\,
_1\phi_1\Bigl(\{0\},\{s^2\},s,s^{2-j}\Bigr)
\label{invert5qpatrick}
\end{equation} 

For $s\gg1$, in
the mode with frequency $\omega_0$, the oscillation
is localized near block $0$, as expected, but as $s$ gets closer to $1$ it
spreads out, as shown in figure~\ref{fig-4}.
Because of the combination of scale invariance and translation invariance,
all the modes are translated versions of the mode with $\omega=\omega_0$
--- they all have the same shape. 
The mode with $\omega=s^{n/2}\omega_0$, is
\begin{equation}
\psi^n_j(s) =\psi_{j-n}(s)\,
\label{psin}
\end{equation}
Thus if the modes are normalized (and taken to be real), $\psi_j(s)$ satisfies
\begin{equation}
\sum_j\psi_j(s)\psi_{j-n}(s)=\delta_{0n}
\label{orthonorm}
\end{equation} 
Summing (\ref{orthonorm}) over $n$ gives another amusing result:
\begin{equation}
\left(\sum_j\psi_j(s)\right)^2=1
\label{amusing}
\end{equation}
We will procede further along this direction in \cite{math}.

As $s$ gets closer to $1$, the modes spread out asymetrically, 
as shown in figure~\ref{fig-5}.
You might guess from the figures that the modes have a smooth continuum
limit as $j$ gets large for $s>0$, and oscillate quickly to zero for
sufficiently negative $j$.  This is more or less correct.  For very large
$j$ and $s$ close to one, the solutions approach Bessel functions.
 To
leading order in $(s-1)$, these are the Bessel functions found in
\cite{Randall:2005me}.
\begin{equation}
\frac{\sqrt{s^{-j}}}{s-1} \left(c_1
   J_1\left(\frac{2
   \sqrt{s^{-j}}}{s-1}\right)+2 i
   c_2 Y_1\left(\frac{2
   \sqrt{s^{-j}}}{s-1}\right)\right
   )
\label{besselrs}
\end{equation}
The more accurate next order result in $(s-1)$ is
\begin{equation}
-\frac{s^{3/2}}{\sqrt{2}} \sqrt{3 s-5}
   \sqrt{s^{-j}} I_1\left(\frac{2
   \sqrt{2} \sqrt{s^{-j}}}{(1-s)
   \sqrt{3 s-5}}\right)
\label{bessel2}
\end{equation}
We will derive this and explore some
subtleties in \cite{math}. 

\section{Gauge couplings\label{gaugecouplings}}

{\figsize\begin{figure}[htb]$$\includegraphics[width=.6\hsize]{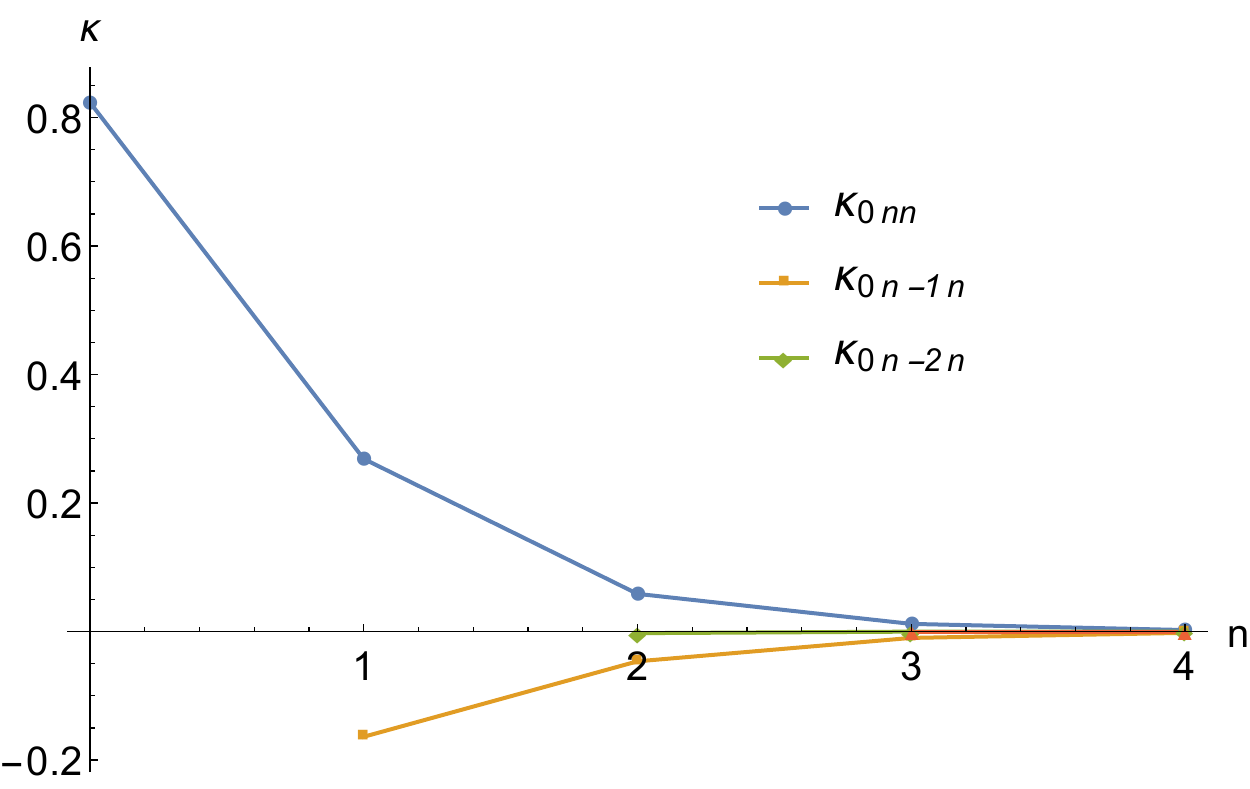}$$
\caption{\figsize\sf
Delocalization of gauge couplings --- Plots of $\kappa_{0n_1n_2}(s)$ for $s=5$. \label{fig-spread5-1}}\end{figure}}
{\figsize\begin{figure}[htb]$$\includegraphics[width=.6\hsize]{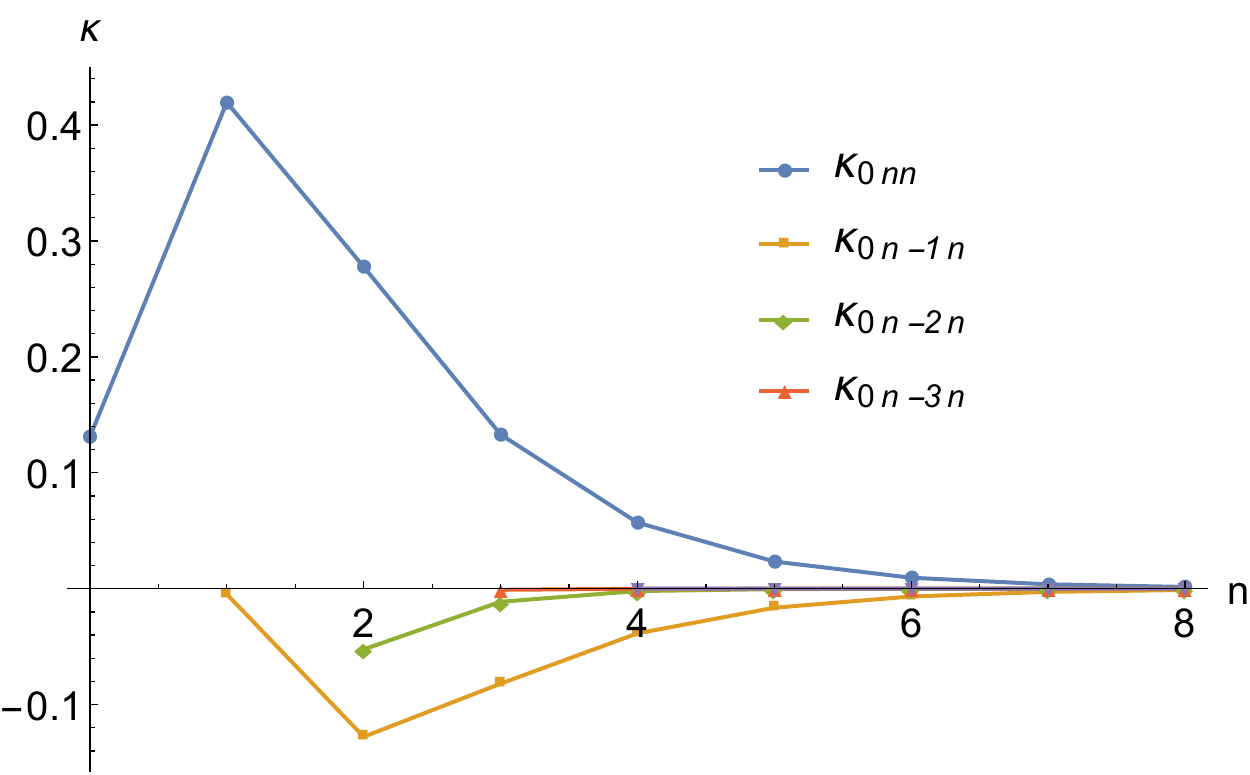}$$
\caption{\figsize\sf
Delocalization of gauge couplings --- Plots of $\kappa_{0n_1n_2}(s)$ for $s=5/2$.\label{fig-spread5-2}}\end{figure}}
{\figsize\begin{figure}[htb]$$\includegraphics[width=.6\hsize]{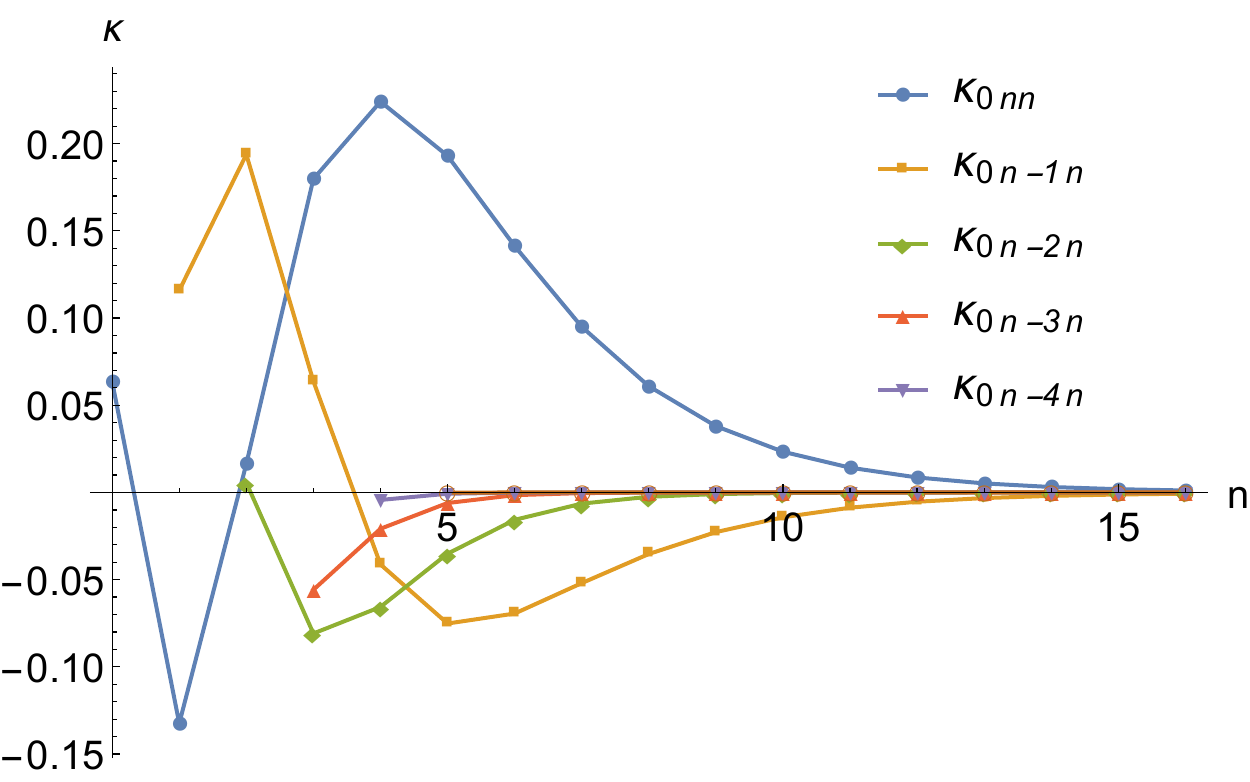}$$
\caption{\figsize\sf
Delocalization of gauge couplings --- Plots of $\kappa_{0n_1n_2}(s)$ for $s=5/3$.\label{fig-spread5-3}}\end{figure}}
In the corresponding quantum field theory, at tree level, the wave
functions of the massive vector bosons are related to the normal modes
(\ref{psin}).  
\begin{equation}
\cl{W}^\mu_n
=\sum_j\psi^n_j(s) W_j^\mu
=\sum_j\psi_{j-n}(s) W_j^\mu
\end{equation}
where $\cl{W}^\mu_n$ is the field with mass $gvs^{n/2}$. 

For large $s$, the state with mass $gv$ has a wave functions
that is localized around $j=0$ --- that is the mass eigenstate gauge field
is approximately $\cl{W}^\mu_0\approx W_0^\mu$ 
with only a small admixture of the other fields.  

For $s$ closer to 1, the wave
functions of the mass eigenstates spread out, like the normal modes.
The trilinear gauge couplings of the mass eigenstates for, $n_1$, $n_2$ and
$n_3$ are proportional to  the symmetric tensor
\begin{equation}
\kappa_{n_1n_2n_3}(s)\equiv \sum_j\psi_{j-n_1}(s)\psi_{j-n_2}(s)\psi_{j-n_3
}(s)
\label{kappa3}
\end{equation}%
Because of the translation invariance of the modes, it is clear that only
differences of the $n$s matter in (\ref{kappa3}).

Empirically, we find that 
the leading contribution to $\kappa_{0jk}$ for $0\leq j\leq k$
and large $s$ has the form
\begin{equation}
\kappa_{0kk}=s^{-k}
\quad\mbox{and}\quad
\kappa_{0jk}=-s^{-(k-j)k}\;\;\mbox{for $j<k$}
\label{scalelarges}
\end{equation}
Note that the largest contribution, $s^{-k}$ in (\ref{scalelarges}) is of
the order of the ratio of the scales between the mass eigenstates involved.

Figures~\ref{fig-spread5-1}, \ref{fig-spread5-2}~and~\ref{fig-spread5-3}
show plots of $\kappa_{0n_1n_2}$.  If $s$ is not too close to $1$, the
couplings are largest when the two largest $n$s are equal and go exponentially
zero if all the $n$s are very different.  
{\figsize\begin{figure}[htb]$$\includegraphics[width=.8\hsize]{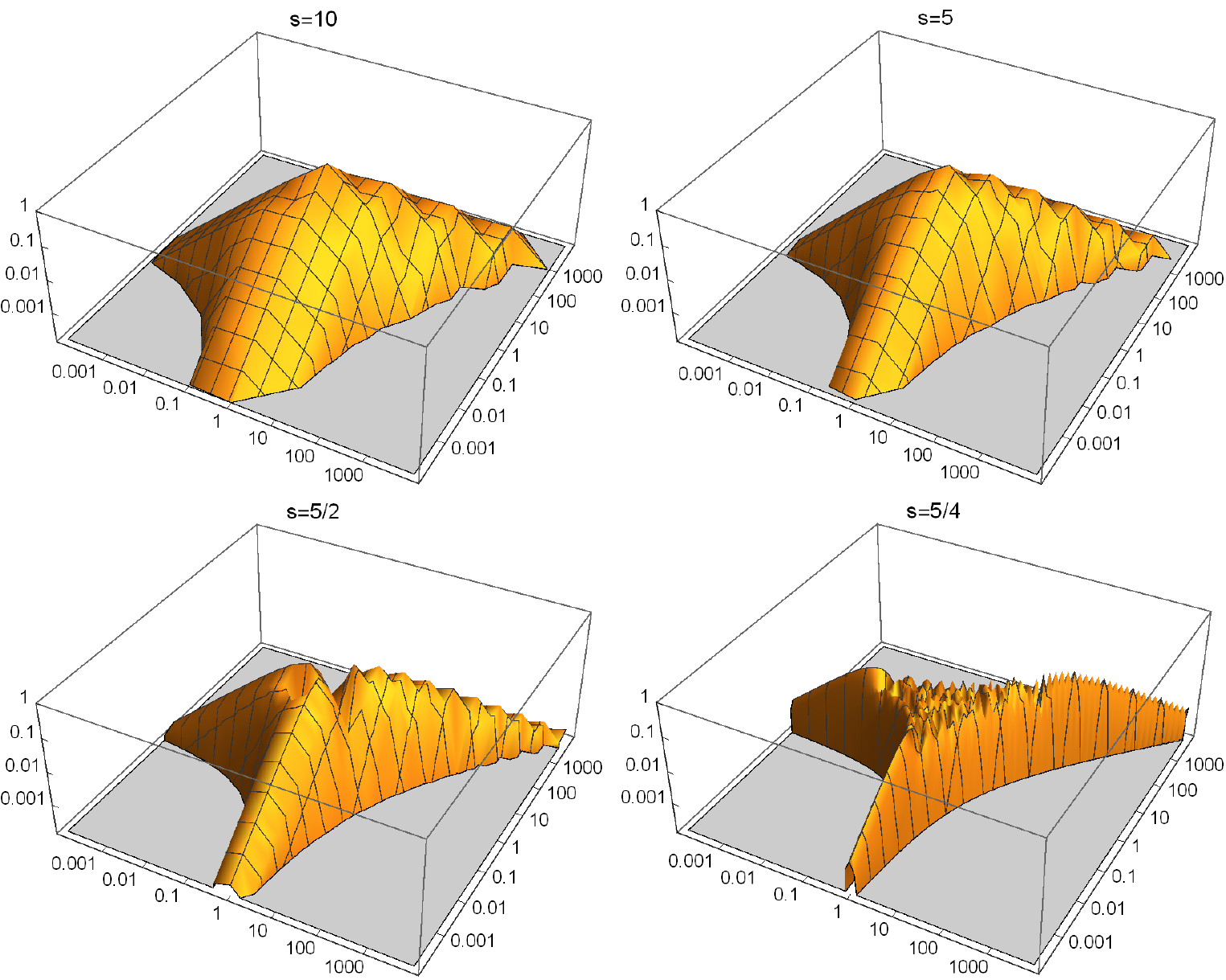}$$
\caption{\figsize\sf\label{fig-spread-grid}$\clog\Bigl(\mathop{\rm
Abs}(\kappa_{0jk})\Bigr)$ versus $s^j$ 
and $s^k$ for various values of $s$ }\end{figure}}

The situation gets quite
complicated for neighboring eigenstates as $s\to1$.
However, if we think about the ``distance'' between the mass eigenstates
not in terms of lattice sites, but in terms of scale, we find that while
spreading does occur as $s$ get's closer to 1, for sufficiently large scale
difference, $\kappa$ still falls like the square of ratio of the small mass to the
large mass, just as in (\ref{scalelarges}).

A nice way to see this is to plot $\clog\Bigl(\mathop{\rm
Abs}(\kappa_{0jk})\Bigr)$ versus $s^j$ 
and $s^k$, the ratios of the mass squares of the $j$ and $k$ eigenstates to
the $0$ eigenstate.  Of course we lose the oscillating fine structure by
taking the absolute value, but it allows us to take the logarithm and
easily see the power dependence for large scale difference.  This
is shown in figure~\ref{fig-spread-grid} for various values of
$s$.\footnote{Note that while the three ridges in the plots look different,
they all actually describe equivalent couplings because
$\kappa_{0,j,j}=\kappa_{0,-j,0}=\kappa_{0,0,-j}$.  The apparent difference
is due to the plot on the rectagular grid which treats the ridge on the diagonal
differently. }

{\figsize\begin{figure}[htb]$$\includegraphics[width=.6\hsize]{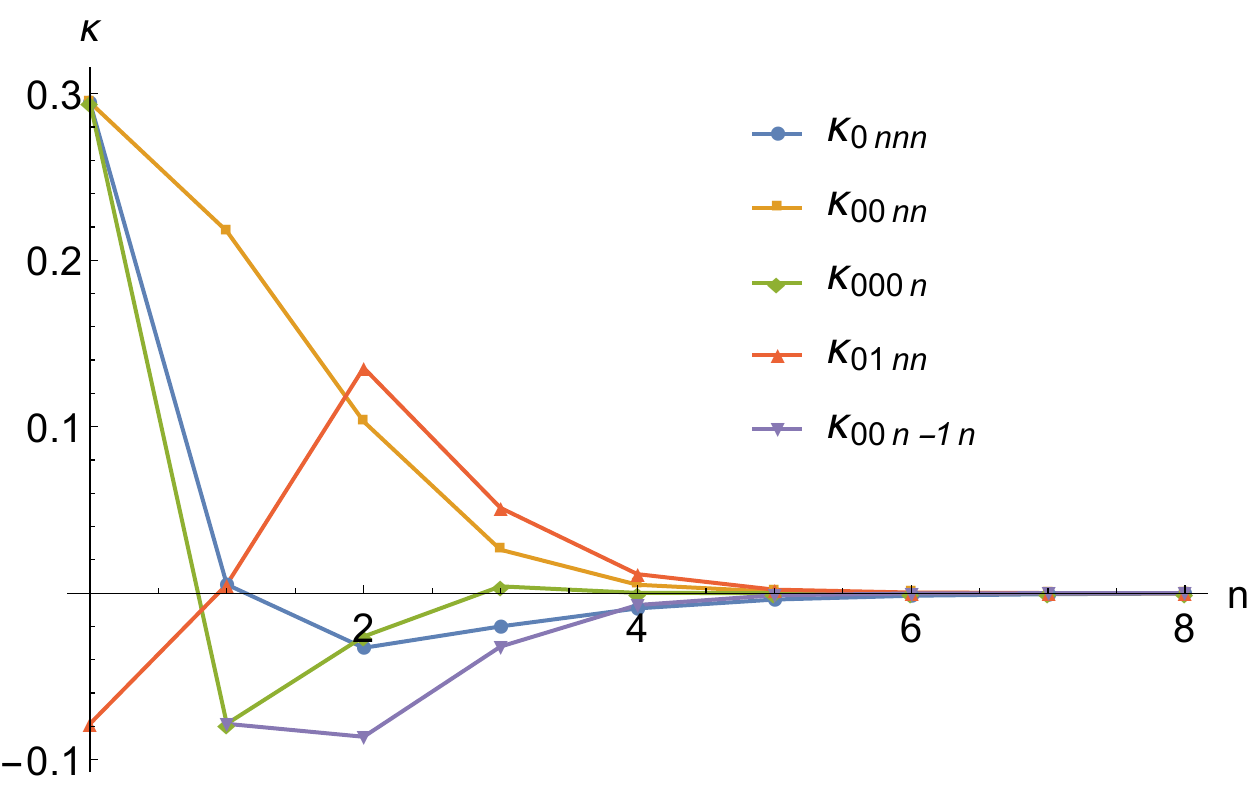}$$
\caption{\figsize\sf
Delocalization of gauge couplings --- Plots of $\kappa_{0n_1n_2n_3}(s)$ for $s=5/2$.\label{fig-spread-4-5-2}}\end{figure}}
Similar delocalization occurs for the quartic couplings, proportional to 
\begin{equation}
\kappa_{n_1n_2n_3n_4}(s)\sum_j\psi_{j-n_1}(s)\psi_{j-n_2}(s)
\psi_{j-n_3}(s)\psi_{j-n_4}(s)
\label{kappa4}
\end{equation}
Figure~{fig-spread-4-5-2} shows some components of (\ref{kappa4}) for
$s=5/2$.  With more indices, the patterns are more difficult to
characterize, but it is clear that the couplings are similarly spread out.

\section{The tale of the missing photon\label{tale}}

{\figsize\begin{figure}[htb]
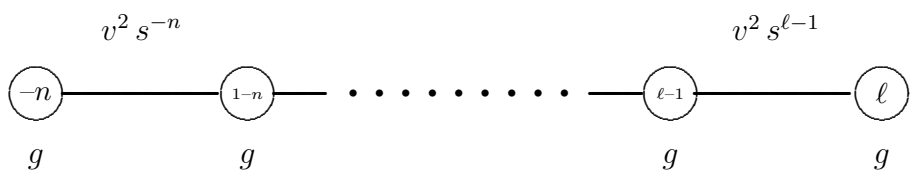

$$\beginpicture
\setcoordinatesystem units <\tdim,\tdim>
\circulararc 360 degrees from 10 0 center at 0 0
\circulararc 360 degrees from 90 0 center at 80 0
\circulararc 360 degrees from 250 0 center at 240 0
\circulararc 360 degrees from 330 0 center at 320 0
\stpltsmbl
\multiput {\tiny$\bullet$} at 120 0 *8 10 0 /
\put {\small--$n$} at 0 0 
\plot 10 0 70 0 /
\put {$v^2\,s^{-n}$} at 40 25
\put {\tiny$1$--$n$} at 80 0
\plot 90 0 110 0 /
\plot 210 0 230 0 /
\plot 250 0 310 0 /
\put {\tiny$\ell$--$1$} at 240 0
\put {\small$\ell$} at 320 0
\put {$v^2\,s^{\ell-1}$} at 280 25
\multiput {$g$} at 0 -25 *1 80 0 /
\multiput {$g$} at 240 -25 *1 80 0 /
\endpicture$$
\caption{\figsize\sf\label{fig-2finite}A truncated, finite version of the system in
figure~\protect\ref{fig-2}.}\end{figure}} 
{\figsize\begin{figure}[htb]
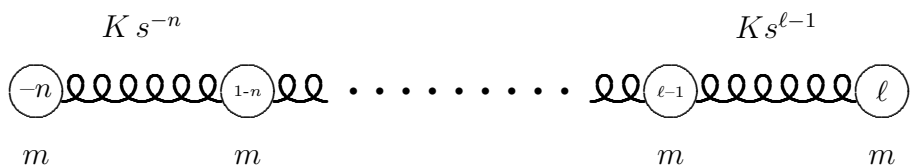

$$\beginpicture
\setcoordinatesystem units <\tdim,\tdim>
\circulararc 360 degrees from 10 0 center at 0 0
\circulararc 360 degrees from 90 0 center at 80 0
\circulararc 360 degrees from 250 0 center at 240 0
\circulararc 360 degrees from 330 0 center at 320 0
\stpltsmbl
\multiput {\tiny$\bullet$} at 120 0 *8 10 0 /
\put {\small--$n$} at 0 0 
\put {\tiny$1$-$n$} at 80 0
\springru 10 -4 *5 /
\put {$K\,s^{-n}$} at 40 25
\springru 90 -4 *1 /
\springru 210 -4 *1 /
\put {\tiny$\ell$--$1$} at 240 0
\put {\small$\ell$} at 320 0
\springru 250 -4 *5 /
\put {$Ks^{\ell-1}$} at 280 25
\multiput {$m$} at 0 -25 *1 80 0 /
\multiput {$m$} at 240 -25 *1 80 0 /
\endpicture$$
\caption{\figsize\sf\label{fig-3finite}The mechanical analog of the finite system in
figure~\protect\ref{fig-2finite}.}\end{figure}} 
If you think about the gauge symmetry of this system naively, you might be
momentarily puzzled by the lack of a massless gauge field.  The vacuum
expectation values 
all preserve the ``diagonal'' gauge symmetry generated by the sum of all
the individual generators.  One might expect massless states corresponding
to the unbroken gauge symmetry.  We will refer to these as ``photons'' for
short.   

But we have seen that there are no massless
states.  The gauge bosons are all massive and associated with the
normalizable normal modes, $\psi^n_j$. These form a complete basis for the
space.  The ``would-be photon'' has the non-normalizable wave 
function 
\begin{equation}
\psi^\gamma_j=1
\label{psigamma}
\end{equation}
which has unit overlap with all the normal modes because of
(\ref{amusing}).  

The physics becomes clear (if it is not already obvious) when you think
about obtaining the infinite systems of figures~\ref{fig-2} and \ref{fig-3}
as the limit of  the finite systems shown in figures~\ref{fig-2finite} and
\ref{fig-3finite}. 
Now, obviously, in the mechanical system, there is a zero frequency mode in
which all the blocks move together with wave function (\ref{psigamma}).
And in the QFT, there is an unbroken gauge symmetry with a 
gauge field proportional to 
\begin{equation}
\sum_{j=-n+1}^{\ell+1}W_j^\mu
\label{gammawf}
\end{equation}
But the gauge coupling is
\begin{equation}
\frac{g}{\sqrt{n+\ell+1}}
\end{equation}
which makes it clear what happens to the photon.  As the system becomes
infinite, the photon mode decouples and becomes unphysical.  
This is consistent with what we see
in the infinite system.

\section{The tail of the missing photon\label{tail}}

{\figsize\begin{figure}[htb]
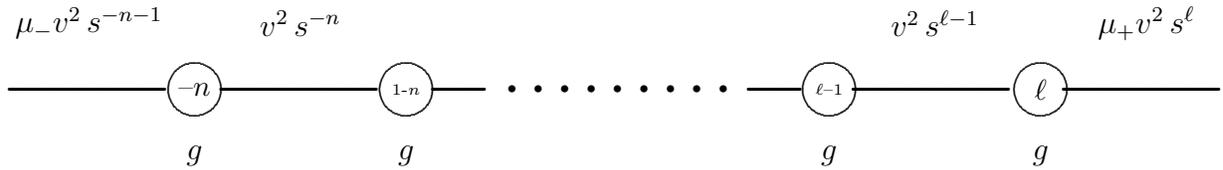

$$\beginpicture
\setcoordinatesystem units <\tdim,\tdim>
\circulararc 360 degrees from 10 0 center at 0 0
\circulararc 360 degrees from 90 0 center at 80 0
\circulararc 360 degrees from 250 0 center at 240 0
\circulararc 360 degrees from 330 0 center at 320 0
\stpltsmbl
\multiput {\tiny$\bullet$} at 120 0 *8 10 0 /
\put {\small--$n$} at 0 0 
\plot 10 0 70 0 /
\put {$v^2\,s^{-n}$} at 40 25
\put {\tiny$1$-$n$} at 80 0
\plot 90 0 110 0 /
\plot 210 0 230 0 /
\plot 250 0 310 0 /
\put {\tiny$\ell$--$1$} at 240 0
\put {\small$\ell$} at 320 0
\put {$v^2\,s^{\ell-1}$} at 280 25
\multiput {$g$} at 0 -25 *1 80 0 /
\multiput {$g$} at 240 -25 *1 80 0 /
\plot -70 0 -10 0 /
\plot 330 0 390 0 /
\put {$\mu_+v^2\,s^{\ell}$} at 360 25
\put {$\mu_-v^2\,s^{-n-1}$} at -40 25
\endpicture$$
\caption{\figsize\sf\label{fig-2boundary}A version of the system in
figure~\protect\ref{fig-2finite} with variable boundary conditions.}\end{figure}} 
{\figsize\begin{figure}[htb]
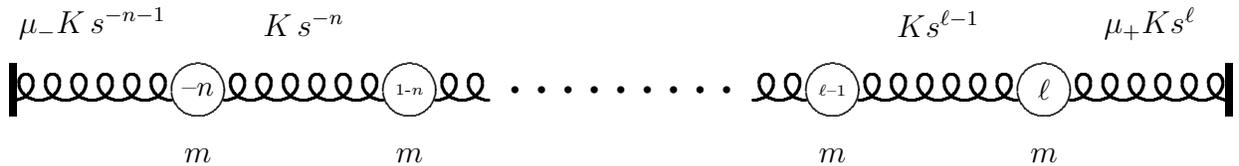

$$\beginpicture
\setcoordinatesystem units <\tdim,\tdim>
\circulararc 360 degrees from 10 0 center at 0 0
\circulararc 360 degrees from 90 0 center at 80 0
\circulararc 360 degrees from 250 0 center at 240 0
\circulararc 360 degrees from 330 0 center at 320 0
\stpltsmbl
\multiput {\tiny$\bullet$} at 120 0 *8 10 0 /
\put {\small--$n$} at 0 0 
\put {\tiny$1$-$n$} at 80 0
\springru 10 -4 *5 /
\put {$K\,s^{-n}$} at 40 25
\springru 90 -4 *1 /
\springru 210 -4 *1 /
\put {\tiny$\ell$--$1$} at 240 0
\put {\small$\ell$} at 320 0
\springru 250 -4 *5 /
\put {$Ks^{\ell-1}$} at 280 25
\multiput {$m$} at 0 -25 *1 80 0 /
\multiput {$m$} at 240 -25 *1 80 0 /
\springru -70 -4 *5 /
\springru 330 -4 *5 /
\linethickness=3pt
\putrule from -70 10 to -70 -10
\putrule from 390 10 to 390 -10
\put {$\mu_-K\,s^{-n-1}$} at -40 25
\put {$\mu_+Ks^{\ell}$} at 360 25
\endpicture$$
\caption{\figsize\sf\label{fig-3boundary}The mechanical analog of the system in
figure~\protect\ref{fig-2boundary}.}\end{figure}} 
The finite systems of figures~\ref{fig-2finite} and \ref{fig-3finite} are
more ``physical'' than the infinite systems of figures~\ref{fig-2} and
\ref{fig-3}.  We can actually build the mechanical system in
figure~\ref{fig-3finite}. And the QFT in figure~\ref{fig-2finite} is more
realistic because it does not assume anything at arbitrarily large and
small energy scales.   On the other hand, the finite systems 
of figures~\ref{fig-2finite} and \ref{fig-3finite} involve very specific
assumptions about how things work at their 
boundaries, and it is worth generalizing this so we can study the very
physical question of how
things depend on boundary conditions.  So in this section, we will study
the systems shown in figures~\ref{fig-2boundary} and \ref{fig-3boundary} in
which the boundary conditions are variable.  The mechanical system of
figure~\ref{fig-3boundary} has springs at the end connected to fixed walls (the
thick vertical lines) and the dimensionless parameters $\mu_\pm$ are the
ratios of the actual spring constants to what we would have in the infinite
system.  The QFT analog figure~\ref{fig-2boundary}
has variable VEVs at the ends  and 
global rather than gauge symmetries at
$j=\ell+1$ and $-n-1$.

{\figsize\begin{figure}[htb]$$\includegraphics[width=.66\hsize]{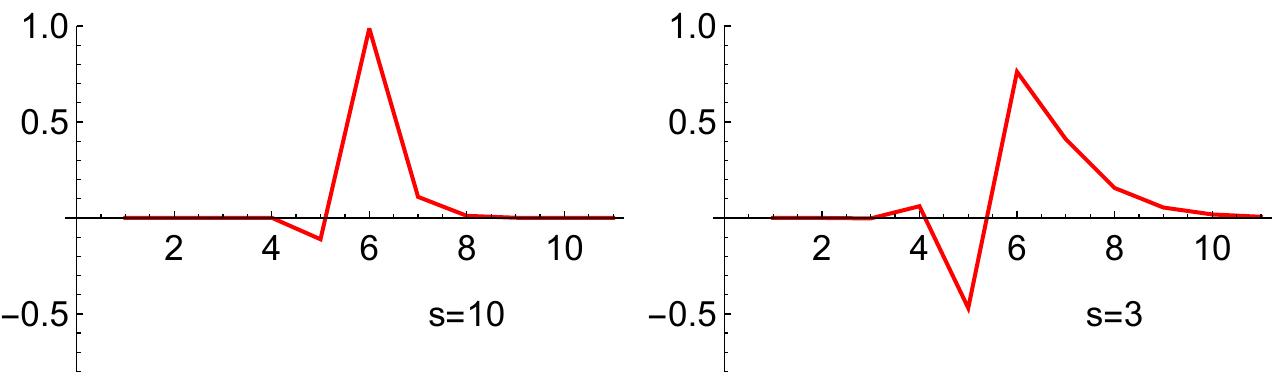}$$
\caption{\figsize\sf
Modes of the infinite system.\label{fig-10and3} }\end{figure}}
We will study the normal modes of figure~\ref{fig-3boundary} as functions
of the boundary parameters, $\mu_\pm$. To begin, let us consider the form
of the normal modes localized in the middle of the system.  What we might
expect from the exponential localization of the modes of the infinite
system is that the effect of the boundary conditions on the central modes
will be exponentially suppressed, of order
\begin{equation}
s^{-(\ell+n)/2}
\end{equation}
{\figsize\begin{figure}[htb]$$\includegraphics[width=.99\hsize]{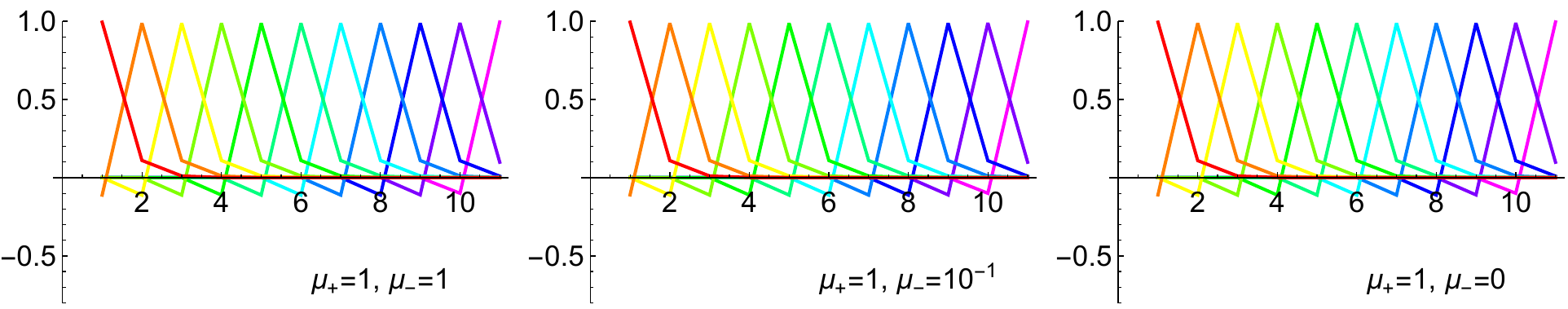}$$
\caption{\figsize\sf
Dependence of the modes of the system in figure~\ref{fig-3boundary} on
$\mu_-$ for $n+\ell=10$, $s=10$ and
$\mu_+=1$.\label{fig-mmm11-10}}\end{figure}} 
{\figsize\begin{figure}[htb]$$\includegraphics[width=.99\hsize]{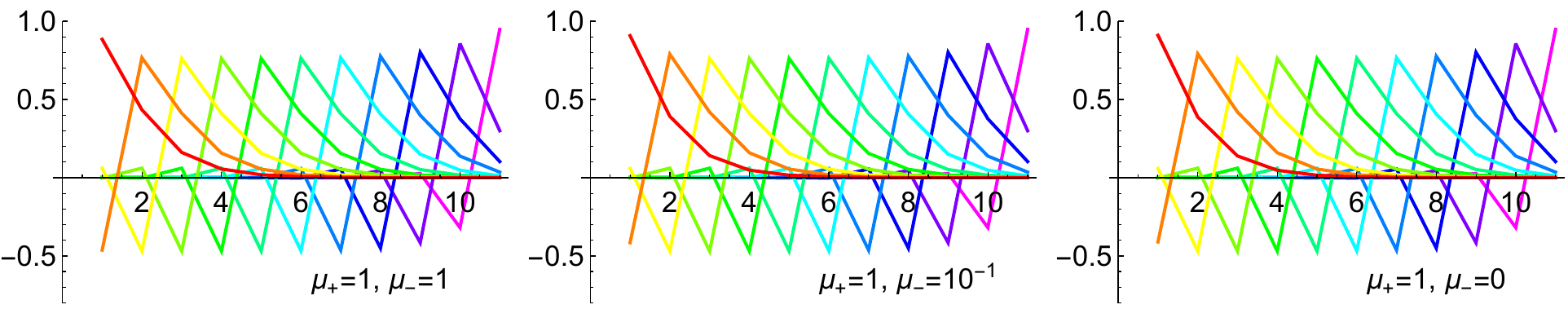}$$
\caption{\figsize\sf
Dependence of the modes of the system in figure~\ref{fig-3boundary} on
$\mu_-$ for $n+\ell=10$, $s=3$ and
$\mu_+=1$.\label{fig-mmm11-3}}\end{figure}} 
And that is indeed how things work for the dependence on $\mu_-$.  We will
illustrate this with three values of $s$, $10$ and $3$ which are
well-enough localized
that they fit comfortably into a finite system with $\ell+n=10$, which is
what we
will use for illustration.

For $\mu_+=1$, you see in figures~\ref{fig-mmm11-10}
and \ref{fig-mmm11-3} that the effect of $\mu_-$ is tiny.\footnote{In these
and subsequent figures, the signs of the individual modes is ambiguous and
I have chosen the signs to make the regularities apparent.}  You {\bf can} see,
particularly in figure~\ref{fig-mmm11-3}, the effect of the boundaries on
the nearby modes.  Obviously, this effect increases as $s$ gets closer to
$1$ and the modes spread out.

{\figsize\begin{figure}[htb]$$\includegraphics[width=.99\hsize]{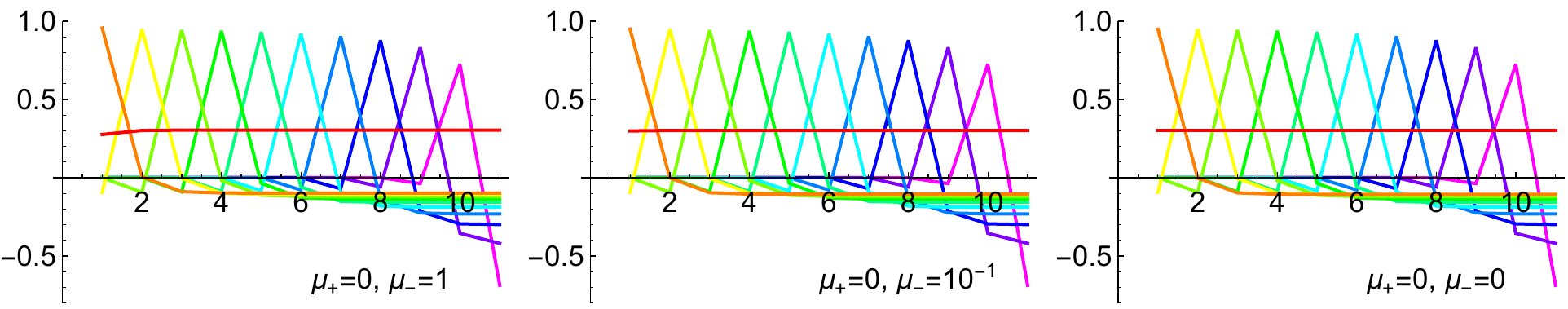}$$
\caption{\figsize\sf
Dependence of the modes of the system in figure~\ref{fig-3boundary} on
$\mu_-$ for $n+\ell=10$, $s=10$ and
$\mu_+=0$.\label{fig-mm11-10}}\end{figure}} 
{\figsize\begin{figure}[htb]$$\includegraphics[width=.99\hsize]{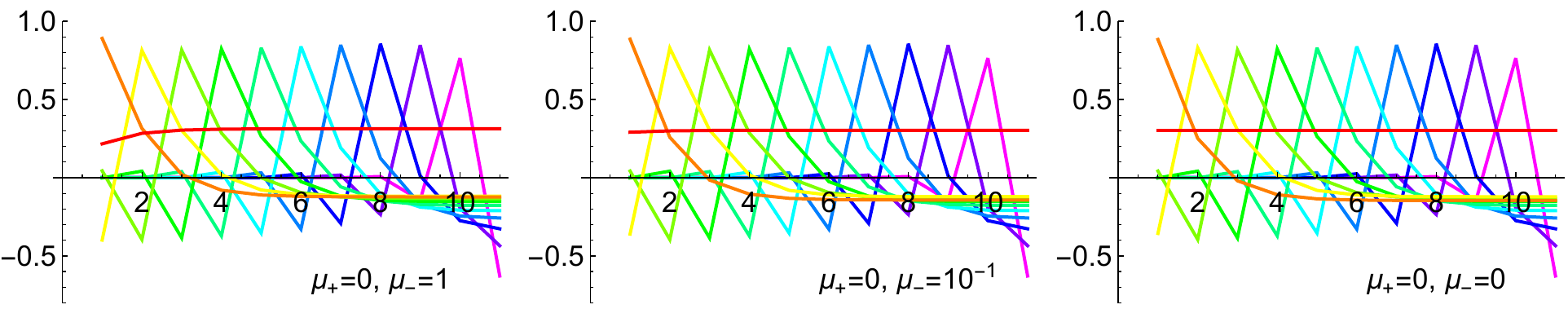}$$
\caption{\figsize\sf
Dependence of the modes of the system in figure~\ref{fig-3boundary} on
$\mu_-$ for $n+\ell=10$, $s=3$ and
$\mu_+=0$.\label{fig-mm11-3}}\end{figure}} 
For $\mu_+=0$, the analogous situation is shown 
in figures~\ref{fig-mm11-10}
and \ref{fig-mm11-3}.  Now there is a photon state for $\mu_-=0$, so a very
small $\mu_-$ can have a very large effect on the lowest mode.  But for the
rest of modes, the dependence on $\mu_-$ is too small to see.  Note
however, that the modes for $\mu_+=0$ look very different from the modes
with $\mu_+=1$.  This is the effect of the photon state (missing or not).  It it worth
pausing here to remark on why it makes such a big difference.  It is helpful to  look at
the form of the modes for $\mu_-=\mu_+=0$ as $s\to\infty$. 
{\figsize\begin{figure}[htb]$$\includegraphics[width=.66\hsize]{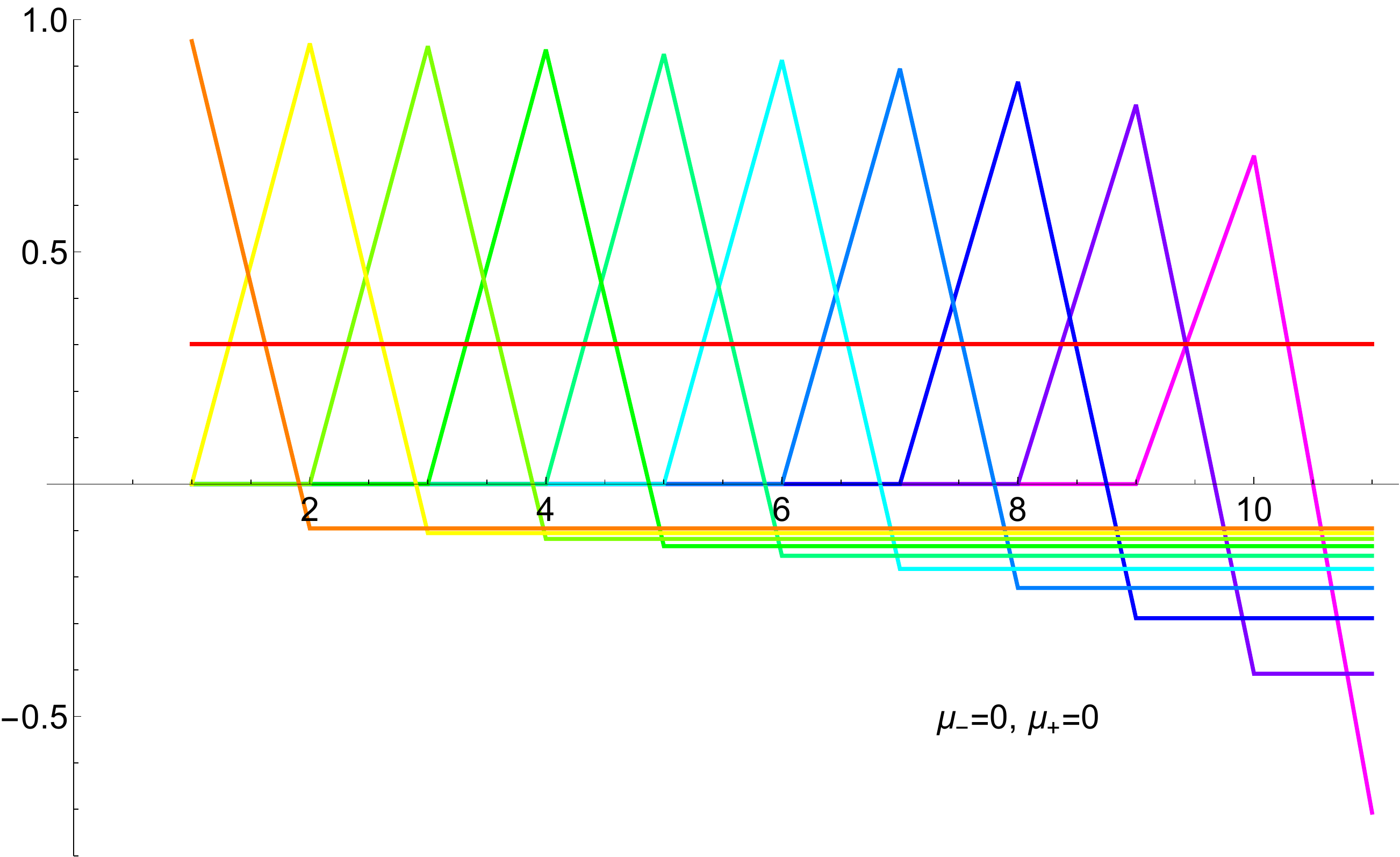}$$
\caption{\figsize\sf
The modes of the system in figure~\ref{fig-3boundary} 
for $\mu_+=\mu_-=0$ for $n+\ell=10$ as $s=\infty$.\label{fig-mm11-infinity}}
\end{figure}} 
In this limit, it is easy to construct the modes by working down from the
highest frequency mode and integrating out modes as we go down.  The result
is that the modes have the form (with $j=1$ the highest frequency mode)
\begin{equation}
A^j_k=
\left\{
{\renewcommand{\arraystretch}{1.3}
\begin{array}{c@{~\mbox{for}~}c}
0&k< n+\ell+1-j\\
\sqrt{\frac{j}{j+1}}&k= n+\ell+1-j\\
-\sqrt{\frac{1}{j(j+1)}}& n+\ell+1-j<k
\end{array}
}
\right.
\label{notphoton}
\end{equation}
for $j=1$ to $n+\ell$ and
\begin{equation}
A^{n+\ell+1}_k=\frac{1}{\sqrt{n+\ell+1}}\quad\mbox{for all $k$}
\end{equation}
where the lowest mode is the photon. 
This limiting form is shown in figure~\ref{fig-mm11-infinity}.
If $n\to\infty$ and the system becomes
semi-infinite, the modes in (\ref{notphoton}) are still valid, and in fact
in this case they are the whole story because the photon mode goes missing
in the limit $n\to\infty$.
But for finite $n$ and $\ell$, (\ref{notphoton}) shows clearly how the
modes manage to remain orthogonal to each other and to the photon mode.
These modes are {\bf mostly} localized about $k=n+\ell+1-j$, but they have
a tail that stretches out to the upper boundary of the system.

{\figsize\begin{figure}[htb]$$\includegraphics[width=.99\hsize]{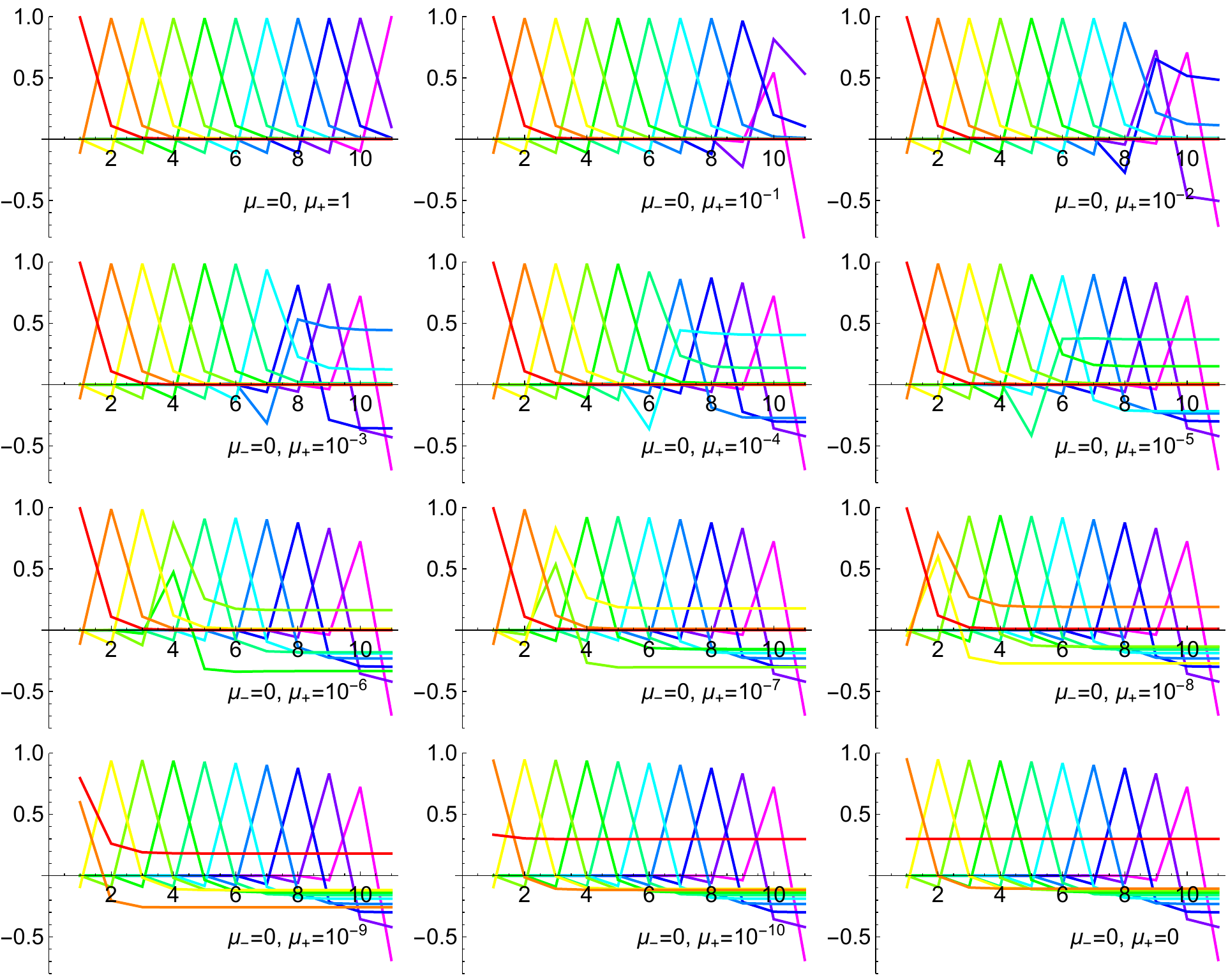}$$
\caption{\figsize\sf
Dependence of the modes of the system in figure~\ref{fig-3boundary} on
$\mu_+$ for $n+\ell=10$, $s=10$ and
$\mu_-=0$.\label{fig-m11-10}}\end{figure}} 
{\figsize\begin{figure}[htb]$$\includegraphics[width=.99\hsize]{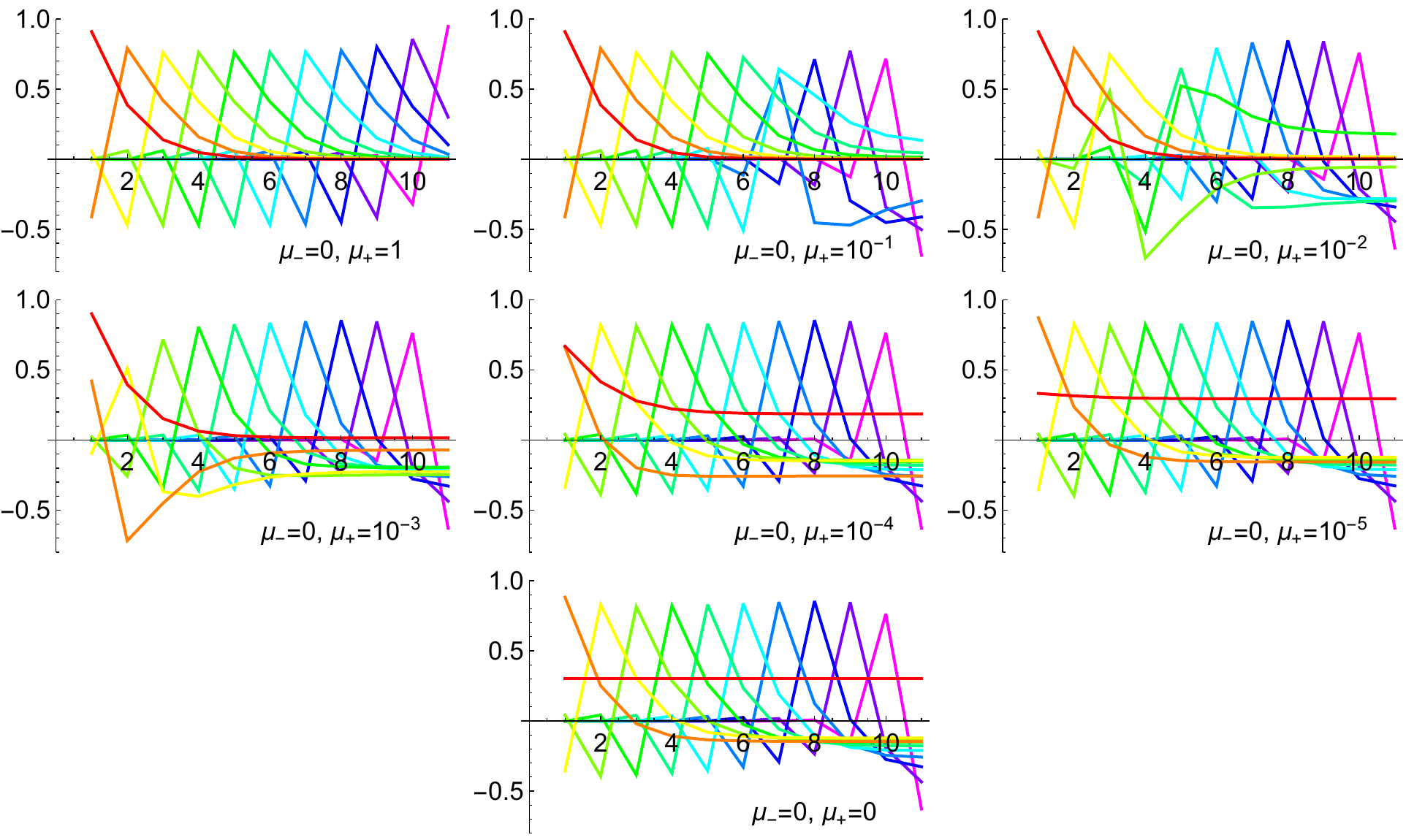}$$
\caption{\figsize\sf
Dependence of the modes of the system in figure~\ref{fig-3boundary} on
$\mu_+$ for $n+\ell=10$, $s=3$ and
$\mu_-=0$.\label{fig-m11-3}}\end{figure}} 
We now turn to the case of $\mu_+$ dependence for $\mu_-=0$.
You can see immediately from figures~\ref{fig-m11-10}
and \ref{fig-m11-3} that something much more interesting is going on in
this case.  
To tease out what is going on, let's begin by plotting the just the lowest
mode for $s=10$.  This is shown in figure~\ref{fig-m11-10-11}. 
{\figsize\begin{figure}[htb]$$\includegraphics[width=.9\hsize]{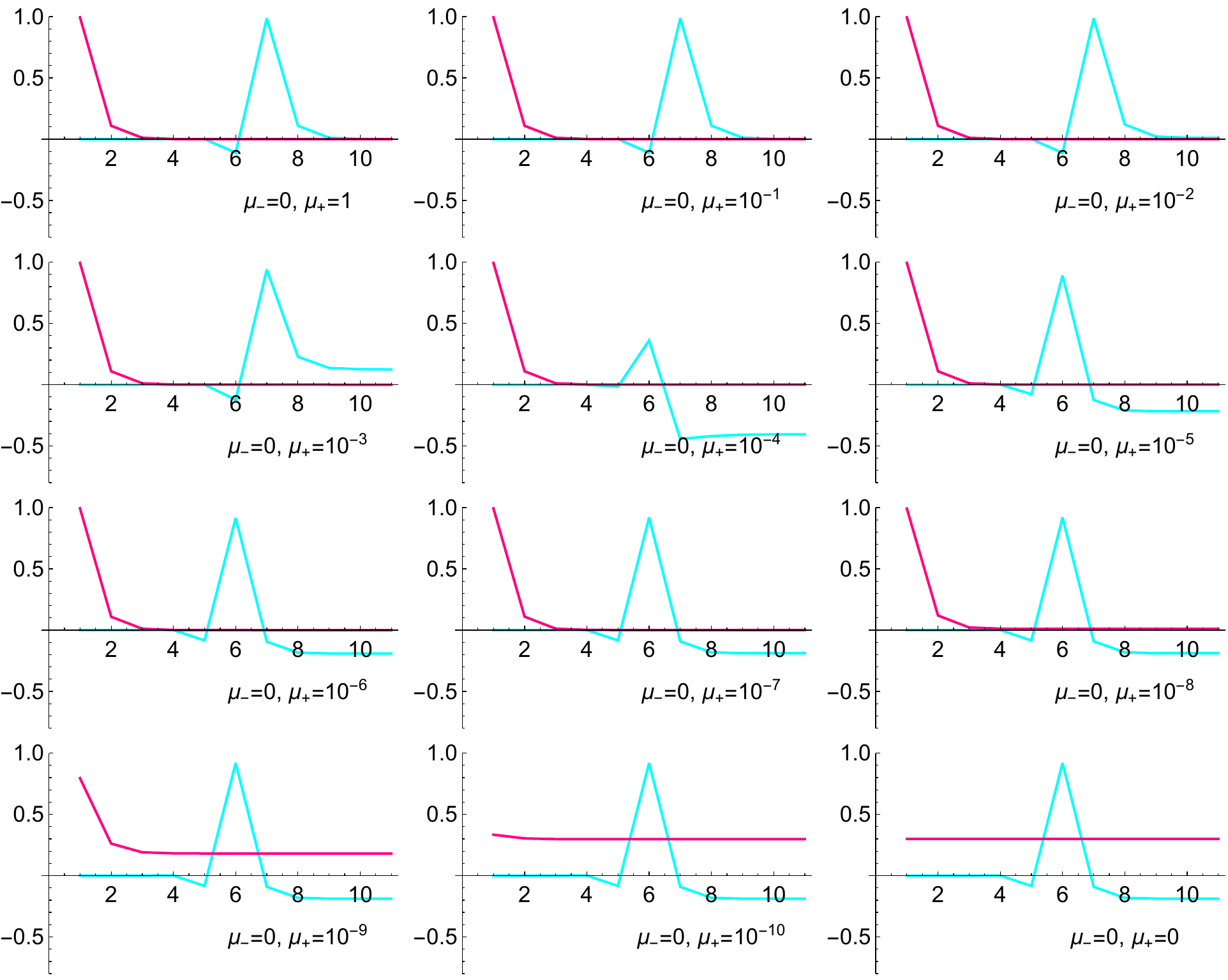}$$
\caption{\figsize\sf\label{fig-m11-10-11}
}\end{figure}} 
{\figsize\begin{figure}[htb]$$\includegraphics[width=.9\hsize]{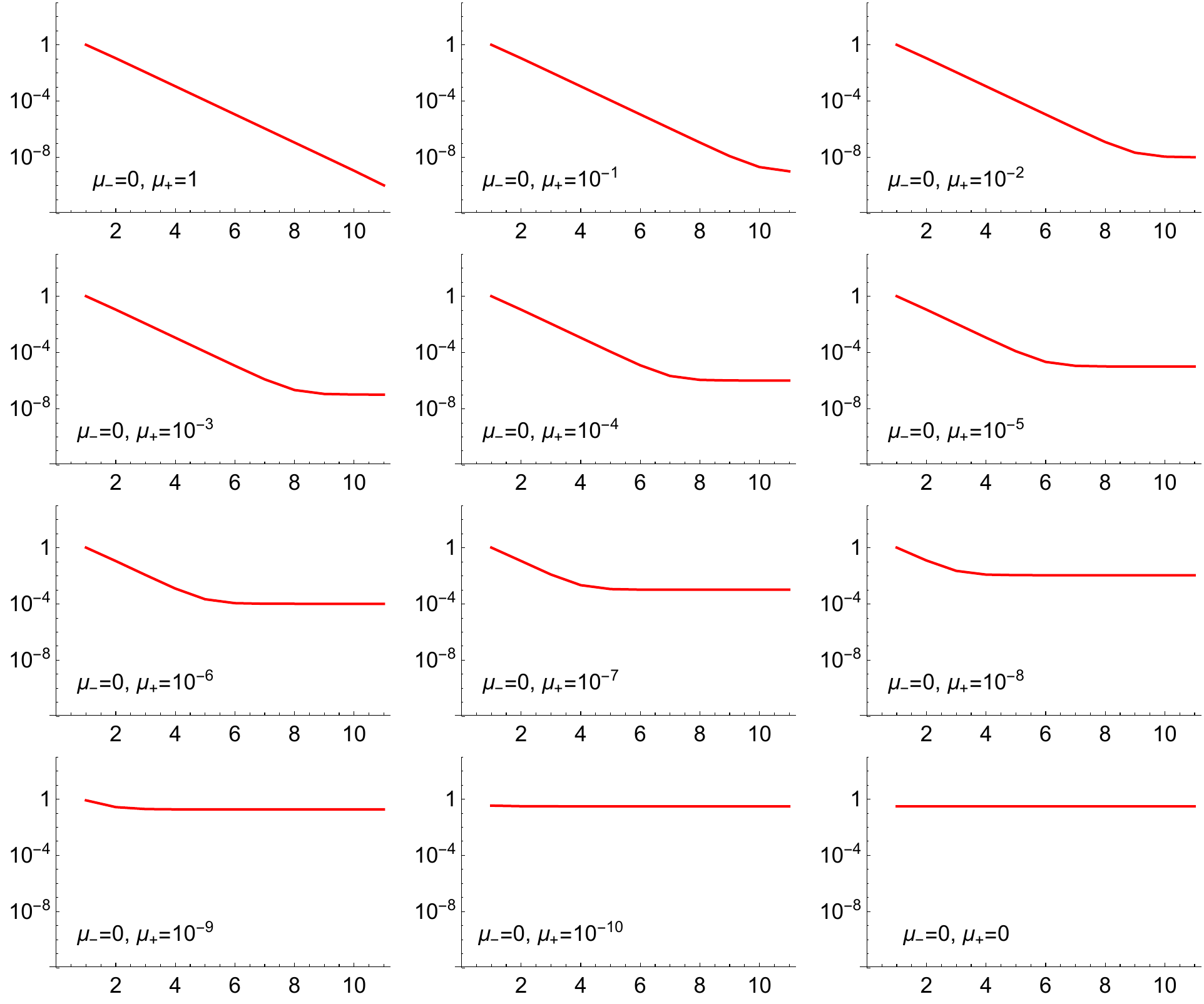}$$
\caption{\figsize\sf\label{fig-m11-10-11-log}
}\end{figure}} 
It is still hard to tell what is going on when the value gets close to
zero, so we can take advantage of the fact that this mode does not change
sign and use a log plot, as shown in figure~\ref{fig-m11-10-11-log}. 

Evidently,
what is happening is that while for $\mu_+=1$, the mode decreases
exponentially for all $j$, for smaller $\mu_+$, the mode decreases out to
about $j=\ell+n+1+\log_{10}\mu_+$, and then it levels out and becomes ``photon
like''.  This break occurs for $j$ such that the local spring constant is
of the order of the spring constant at the upper boundary.  For larger $j$,
the effect of the boundary spring is negligible compared to the local
springs and the rest of the mode looks photon-like.  In this way, the
lowest mode changes smoothly from the exponentially decreasing mode at
$\mu_+=1$ to a photon-like constant mode for $\mu_+<10^{-10}$.

For general $s$, the break occurs for $j\approx\ell+n+1+\log_{s}\mu_+$. For
larger $j$ (smaller frequencies), the modes are localized and resemble the
corresponding modes 
in figure~\ref{fig-mmm11-10} and \ref{fig-mmm11-3}.  For smaller $j$
(higher frequencies) the modes have a tail out to the right boundary and resemble the
corresponding modes 
in figure~\ref{fig-mm11-10} and \ref{fig-mm11-3}.

\section{Where's the physics?\label{physics}}

In the infinite system, we have equivalent physics at every scale because
we have built in scale invariance.   And the physics is local in scale.
Objects of a certain mass interact preferentially with objects of similar
mass.  We have seen in section~\ref{gaugecouplings} that the gauge
couplings between heavy gauge fields with very different masses are
power-law suppressed.  This is what we might expect from effective field
theory.  Even as $s\to1$, while the objects themselves spread out, the
locality in scale persists at sufficiently large scales.  

But the dependence on boundary conditions of the wave functions that we
identified in section~\ref{tail} is quite different.  We have seen
in the figures that the details of the boundary condition at
the high scale has a dramatic effect on the structure of the low-energy end
of the chain.  The very existence of the photon is an obvious example, but
in addition, the long tails of the other low-lying modes in
figure~\ref{fig-mm11-infinity}
and the strong
dependence of the low-lying modes on a small VEV at the large scale shown in
figure~\ref{fig-m11-10} and \ref{fig-m11-3}.
All of these effects are in some sense small.  For the low lying modes,
they are suppressed by a factor of the order of
\begin{equation}
\frac{1}{n+\ell}
\label{linear}
\end{equation}
But because the inverse of 
this factor is {\bf linear} in the distance between in mode in
term of number of sites, it is {\bf logarithmic} in the ratio of large scale over the
small scale.  Thus these are effects of order
\begin{equation}
\frac{1}{n+\ell}
=
\frac{\mathop{\rm Log(s)}}{\clog\Bigl(s^\ell/s^{-n}\Bigr)}
=
\frac{\mathop{\rm Log(s)}}{\clog\Bigl(m_{\rm large}^2/m_{\rm small}^2\Bigr)}
\label{logarithmic}
\end{equation}

I do not have any firm conclusions from this exercise, but I find
(\ref{logarithmic}) rather tantalyzing.  It looks like a violation of
locality in scale that depends critically on the discreteness of the scale
invariance in the model, because the effects are proportional to $\clog(s)$
and disappear as $s\to1$.    I do not know how to construct models of this
kind in a natural way, but I think that any sort of violation of standard
effective field theory power counting is worth exploring further.  

\section*{Acknowledgements}

Matt Schwartz and Patrick Kamiske participated in early stages of this
exploration and I am very grateful for many important discussions with them.
Matt Reece was extremely helpful in pointing me to the literature on
deconstructed AdS(5) and it is not his fault if I have left out important
connections to this large body of work. I am also grateful to Jesse Thaler
for suggestions.
HG is supported in part by the National
Science Foundation under grant PHY-1418114. 

\bibliography{funwithdsi}

\end{document}